\documentclass{PoS}

\title{Chiral symmetry breaking in lattice brane QED model}

\ShortTitle{Chiral symmetry breaking in lattice brane QED model}

\author{\speaker{Eigo Shintani}\\
        RIKEN-BNL Research Center, Brookhaven National Laboratory, Upton, NY 11973, USA\\
        E-mail: \email{shintani@riken.jp}}

\author{Tetsuya Onogi\\
        Department of Physics, Osaka University, Toyonaka 560-0043, Japan\\
        E-mail: \email{onogi@het.phys.sci.osaka-u.ac.jp}}

\abstract{
We propose a novel lattice calculation of spontaneous chiral symmetry breaking 
in QED model with 2+1 dimensional fermion brane.
Considering the relativistic action with gauge symmetry 
we rigorously carry out path integral in Monte-Carlo simulation
with Fermi-velocity relevant to effective coupling constant.
We numerically show the evidence of spontaneous chiral symmetry breaking 
in strong coupling region with chiral condensate, low-lying mode distribution and 
Nambu-Goldstone boson spectrum in bare Fermi-velocty $v=0.1$.
This is a feasible study to investigate the phase structure of Graphene. 
}
\FullConference{The 30 International Symposium on Lattice Field Theory - Lattice 2012,\\
		June 24-29, 2012\\
		Cairns, Australia}
\begin{document}

\section{Motivation and background}
We consider QED model with 2+1 dimensional fermion brane (bQED$_3$) 
\cite{Gorbar:2001,Shintani:2012sz}; 
in this model fermion has smaller velocity than speed-of-light $c$ 
whose magnitude has been roughly estimated as $c/300$ 
\cite{Wallace:1947}. 
Tight-binding approximation of Graphene \cite{Novoselov:2004}
(or similar material formed as honeycomb seat of atoms)
has suggested that will possess a particular property 
which is described by a
massless Dirac particle (quasiparticle)
\cite{Wallace:1947,Semenoff:1984}.
Although there are many analytic studies 
\cite{Vafek:2007,Son:2007,Gamayun:2010,Juricic:2009}
and numerical works \cite{Drut:2009,Armour:2010}, 
it has not been successful 
to clearly describe the electric property of Graphene. 

Many models argue that there should be a signature of 
spontaneous gap generation due to strong dynamics
which indicates the occurrence 
of semiconductor-insulator transition for suspended monolayer Graphene.
However in the experiment there is no evidence of the 
gap of band structure between electron-hole, and thus
it turns out to be permanently conductor.
Such inconsistency may be due to the suppression effect of silicon bases,
whose dielectric permittivity $\varepsilon$ appears
in the denominator of effective coupling, 
$\alpha_e(\varepsilon) = 2\alpha_e / (1+\varepsilon) \ll 1$, and 
thus the impurity of Graphene might be cause of no phase transition. 
To verify this argument the identification of critical point of effective
coupling constant from theoretical study is needed.
Our motivation is clarification of phase structure of bQED$_3$ model
to understand the transition of Graphene and other kinds of 
honeycomb crystal.

\section{Brane QED$_3$ model}




As pointed out in \cite{Shintani:2012sz} the {\it relativistic} bQED$_3$ model 
enables us to rigorously deal with path integral under gauge invariant formalism,
\begin{equation}
  S_{\rm bQED_3} = \frac{\beta}{2}\int dtdx^3\Big(v\vec E^2 + v^{-1}\vec B^2\Big)
  + \int dtdx^2\bar\psi\Big[iD_t\gamma_t + iv(D_x\gamma_x + D_y\gamma_y)\Big]\psi
\label{eq:bQED_3}
\end{equation}
with {\it bare} Fermi-velocity parameter. 
In this action we consider not only the Coulomb interactions but also induced 
magnetic interactions which are $v^2$ times weaker than electric field. 
Since we attempt to treat bQED$_3$ model as local field theory 
(super-renormalizable), 
the theoretical parameters, $v$ and $\beta$, are affected by renormalization. 
The argument of perturbation theory in bQED$_3$ model is that 
the Fermi-velocity has a logarithmic divergence.
The experiment \cite{Elias:2011} supports 
the renormalization effect of fermi-velocity of Graphene
as suggested in perturbative bQED$_3$ model \cite{Gonzalez:1994,Kotov:2010}
rather than non-relativistic one 
\cite{Vafek:2007,Son:2007,Gamayun:2010,Juricic:2009,Drut:2009,Armour:2010}.
The effective coupling constant is defined as the modified form as 
$\alpha_e = 1/(4\pi\beta v)$
and 2-loop analysis in approximated perturbation explicitly shows 
the UV fixed point in $\alpha_e \gg 1$ \cite{Kotov:2010}. 
The check of existence of UV fixed point with non-perturbative method 
is interested in this work.

The gap generation of Graphene is considered to be related with 
spontaneous chiral symmetry breaking ($\chi$SB) in bQED$_3$ model. 
This is analogous to the second order chiral phase transition 
in $N_f=2$ massless QED$_3$ involving 
the mass gap \cite{Pisarski:1984dj,Appelquist:1986}. 
In Eq.(\ref{eq:bQED_3}) quasiparticle field $\psi$ has 4 spinor component, 
$\psi^t = (\psi_\sigma^{A+},\psi_\sigma^{B+},\psi_\sigma^{B-},\psi_\sigma^{A-})$
which corresponds to three kinds of Graphene symmetry; 
Dirac valleys (degenerating ground energy) ($\pm$), 
sublattice symmetry $(A,B)$ and spin of carbon atoms $(\sigma)$.
The gamma matrix is given as the tensor structure; 
$\gamma_t = \sigma_0\otimes I_{2\times2}$, 
$\gamma_i = -i\sigma_2\otimes \sigma_i (i=x,y,z)$
in which the first part is degree of freedom (DOF) in valley times sublattice 
and the second one is DOF in spin rotation.
Regarding the Graphene symmetry as "flavor" U(4) symmetry in bQED$_3$, 
whose 16 generators are represented as 
$\{1,\gamma_5,i\gamma_3,[\gamma_3,\gamma_5]/2\}\otimes \sigma_{i=t,x,y,z}$ with
$\gamma_5 = i\gamma_t\gamma_x\gamma_y\gamma_z$, 
we define $\gamma_5 = {\rm diag}(1,1,-1,-1)$ "chiral" projection corresponding 
to left and right chirality (valley chirality) as $U_L(2)\times U_R(2)$.
The chiral condensate $\langle\bar\psi\psi\rangle$ is used to be 
the order parameter of $\gamma_5$ $\chi$SB as well as QCD.

\section{Lattice calculation of brane QED$_3$ model}

In order to carry out {\it ab initio} calculation of bQED$_3$ model, 
we implement the staggered-type fermion action including the 
non-compact U(1) gauge action;
\begin{eqnarray}
  S_g &=& \sum_{n=(x,y,z,t)} \Big[ \beta v\sum_i(\nabla_4\theta_i(n)-\nabla_i\theta_4(n))
  + \beta v^{-1}\sum_{i,j}(\nabla_i\theta_j(n)-\nabla_j\theta_i(n))\Big],\\
  S_f &=& \sum_{m=(x,y,t)}\Big[ \sum_i \eta_i(m)\bar\chi(m)\Big\{U_i(m)\chi(m+\hat i)
         - U^\dag_i(m)\chi(m-\hat i)\big\} + M\bar\chi(m)\chi(m)\Big],
\end{eqnarray}
with differential $\nabla_i(m,n) = \delta_{m,n+1}-\delta_{m,n}$,
link variable $U_i(m) = \exp(i\theta(m))$ and Kawamoto-Smit phase factor 
$\eta_i(m) = \prod_{k=1}^{m_{i-1}}(-1)^{k}$.
Conveniently the staggered action in 2+1 dimension 
has $U(2)\times U(2)$ (flavor-) chiral symmetry in the continuum limit,
and hence we can perform Hybrid-Monte-Carlo (HMC) simulation 
straightforwardly \cite{Drut:2009,Armour:2010,Shintani:2012sz}.

Here we show the numerical results of chiral condensate, low-lying mode distribution 
and hadronic spectrum in $N_s\times N_t = 40^2\times 20$ and $N_z=8$ lattice 
with fixed $v=0.1$ as a function of fermion mass $M$ and coupling constant $\beta$. 
For U(1) field we use periodic boundary for spatial and temporal directions, 
and for fermion field is periodic boundary for spatial direction 
and anti-periodic boundary for temporal one.
This simulation realizes the lower temperature system
thanks to a rescaled temporal size by velocity.
To avoid the autocorrelation we use every 20 HMC trajectory per configuration, 
and the error analysis adopts the Jackknife(JK) method with 10 bin size.
Chiral condensate is estimated by 100 noise sources in each configurations.

\section{Chiral symmetry breaking in brane QED$_3$ model}
\subsection{Chiral condensate}

\begin{figure}
\begin{center}
\includegraphics[width=90mm]{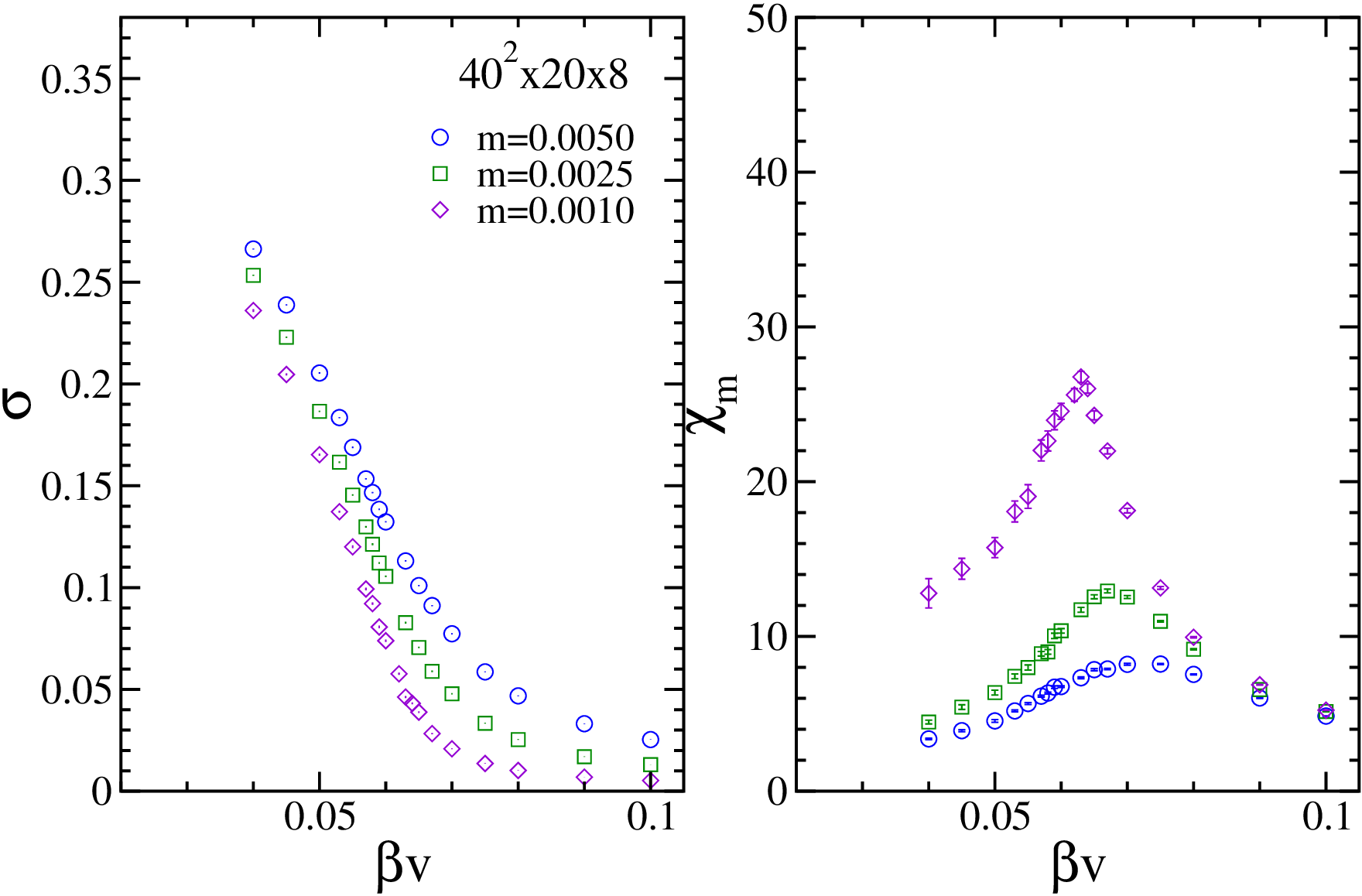}
\caption{The dependence of the inverse of effective coupling constant
$1/(\beta v)$ for chiral condensate $\sigma$ (left) 
and chiral susceptibility $\chi_m$ (right).
Different symbols denote the points in the different fermion masses.}
\label{fig:sigma}
\end{center}
\end{figure}

We first show the chiral condensate and chiral susceptibility;
\begin{equation}
 \sigma\equiv \langle\bar\chi\chi\rangle 
 = \Big\langle \sum_m D^{-1}(m,m)\Big\rangle,\quad
 \chi_m\equiv \frac{\partial\sigma}{\partial M}
   = \Big\langle \Big(\sum_mD^{-1}(m,m)\Big)^2
     -\frac{1}{V}\sum_{m} |D^{-1}(m,0)|^2\Big\rangle
\end{equation}
in several $\beta$ and $M$, where $D^{-1}$ denotes the inverse of 
staggered fermion matrix.
Figure \ref{fig:sigma} shows $\sigma$ and $\chi_m$ have clear dependence of both 
bare effective coupling $\alpha = 1/(4\pi\beta v)$ and mass $M$.
$\sigma$ drastically grows up at $\beta v \simeq 0.05$ -- 0.06, and 
decreasing $M$ this growth is further developing. 
It seems that the critical point exists around $\beta v \simeq 0.05$ -- 0.06
which corresponds to $\alpha_c \simeq 1.3$ -- 1.6, and 
in $\alpha<\alpha_c$ $\sigma$ is close to zero while in $\alpha>\alpha_c$ 
$\sigma$ remains in finite value. 
The chiral susceptibility is clearly shown to be significant
$M$ and $1/(\beta v)$ dependence similar to critical transition.
In $\alpha_c\simeq 1.3$ -- 1.6 there will be a singular point for $\chi_m$ 
in $M=0$ limit, which indicates that $\alpha_c$ 
is expected to be discontinuous at the critical point.

\subsection{Low-lying mode distribution}

Investigation of low-lying mode of massless Dirac operator (Dirac kernel) is helpful to
quantify renormalized chiral condensate $\Sigma$ through the 
Banks-Casher relation;
\begin{eqnarray}
  \Sigma/\pi = \lim_{\varepsilon\rightarrow 0}
               \lim_{V\rightarrow \infty}\rho(\varepsilon),\quad
  \rho(\lambda) = \frac{1}{V}\Big\langle\sum_n\delta(\lambda-\lambda_n)\Big\rangle,
  \label{eq:rho}
\end{eqnarray}
in chiral broken phase.
Spectral density $\rho(\lambda)$ is as a function of 
the eigenvalue $\lambda_n$ of Dirac kernel (here we consider Dirac 
kernel in finite volume $V$, and thus distribution of $\lambda$ 
is a discretized distribution.).
The Banks-Casher relation expects that spectral density shows 
a constant distribution near zero eigenvalue.
In finite volume, according to the discussion of random matrix theory (RMT),
the ``hard edge'' \cite{Damgaard:2000cx} of $\rho$ appears at $\lambda=0$ with
width $1/(V\Sigma)$. 
On the other hand in symmetric phase the spectral density $\rho(\lambda)\sim \lambda^{d-1}$ 
($d$ denotes dimension of fermion), at weak coupling regime. 
This different behavior is used to not only distinguish the phase 
in effective coupling constant as well as Figure \ref{fig:sigma} 
but also quantitatively estimate the $\Sigma$ in 
constant region of $\rho$ near $\lambda\sim 0$. 
In the calculation of low-lying eigenvalue we evaluate 75 different eigenpairs 
(staggered-type Dirac kernel has a pair of eigenvalue with different sign).

\begin{figure}
\begin{center}
\includegraphics[width=90mm]{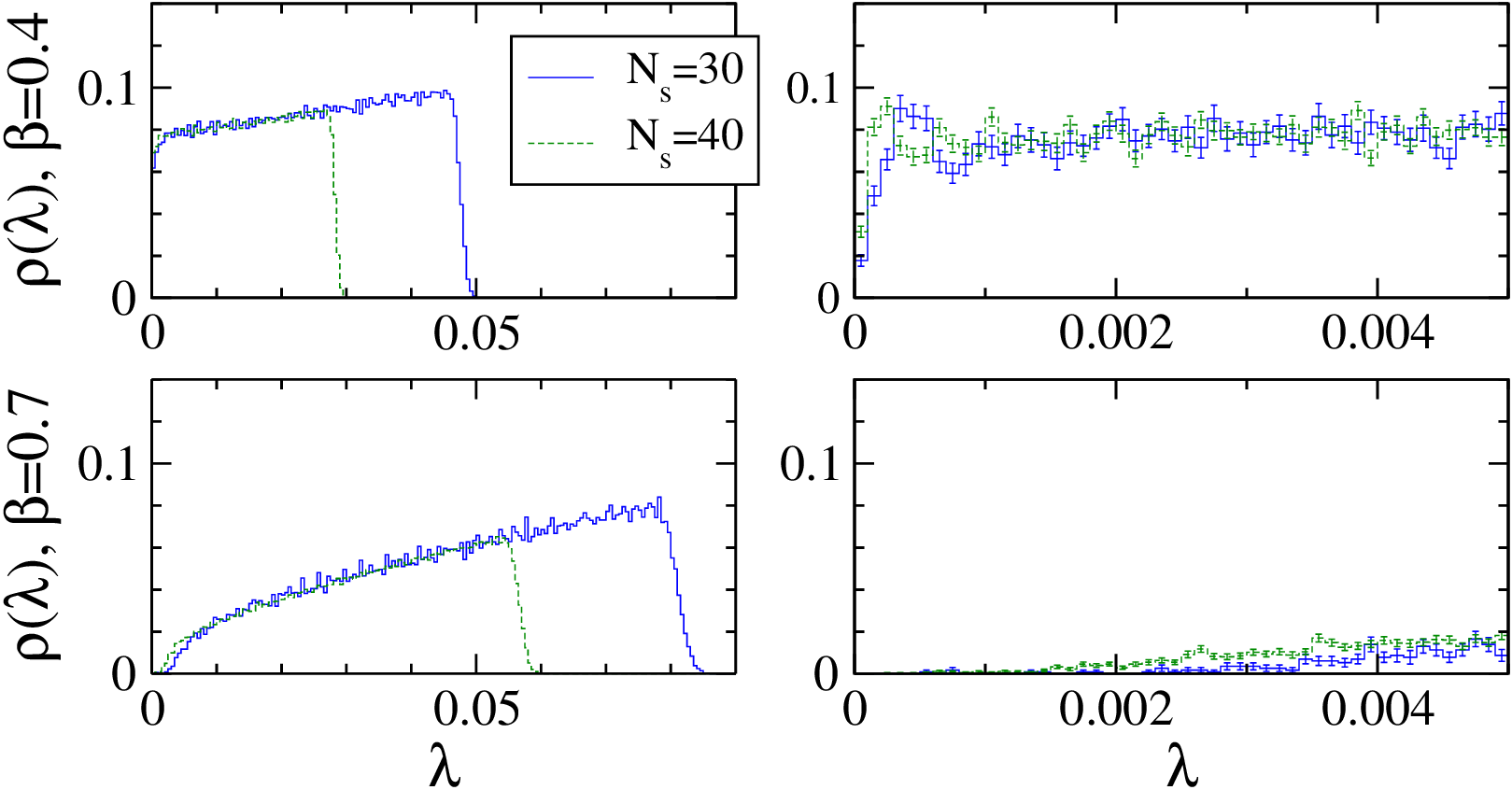}
\caption{Left panel shows the spectral density 
at different $\beta$ which are above, close and 
below $\alpha_c$. We also compare different spatial volume.
Right panel is a zoomed-up right panel near $\lambda=0$.
This is result in sea fermion mass $M_{\rm sea}=0.001$.}
\label{fig:rho}
\end{center}
\end{figure}

In Figure \ref{fig:rho} we can clearly see a significant change of the 
low-lying distribution of $\rho$ above and below $\alpha_c$ which is observed 
in Figure \ref{fig:sigma}.
Below $\alpha_c$ ($\beta \simeq$0.4), there is clear plateau starting from 
$\lambda = 0.002$ in which spectral density is consistent with one in 
different volume. 
Near zero point there is also expected ``hard edge'' whose width 
becomes narrow increasing volume as in the chiral broken phase.
When decreasing the $\alpha$, this ``hard edge'' disappears and 
in $\alpha<\alpha_c$ $\rho$ becomes monotonically growing. 
Fitting in $0.002\le\lambda\le 0.007$ with constant function, 
we obtain $\Sigma$ in each sea fermion mass. 

\subsection{Nambu-Goldstone boson spectrum}

In order to confirm $\Sigma$ obtained in spectral density, we attempt to
reproduce this from measurement of the spectrum of 
Nambu-Goldstone (NG) boson particle in bQED$_3$. 
In NG theorem, when spontaneous $\chi$SB occurs,
NG bosons appear as an asymptotic one-particle state 
coupled with axial-vector current 
(or pseudoscalar via Ward-Takahashi identity) operator. 
If $\alpha>\alpha_c$ is in the chiral broken phase, four NG bosons 
exist as well as pion (and eta) meson in QCD (note that in bQED$_3$ model,
since there is no anomaly, flavor singlet particle can be regarded as 
massless NG boson.)
If this picture is true, we can extract the NG boson amplitude 
and mass from pseudoscalar (PS) correlator in large distance separation
between source and sink point.
After taking large separation of $x$ direction we expect that 
PS correlator which is defined as the exact NG boson operator 
in staggered-type fermion approaches to the following form:
\begin{equation}
  \lim_{x\gg 1} G_{\rm PS}(x) 
  = |Z_{NG}|^2( e^{-m_{\rm NG}x} +  e^{-m_{\rm NG}(N_s-x)})/(2m_{\rm NG}),\quad
  f_{\rm NG} = 2MZ_{\rm PS} m_{\rm NG}^{-3/2},
  \label{eq:G_PS}
\end{equation}
where $f_{\rm NG}$ is a corresponding quantity to pion decay constant in QCD
although its mass dimension is 1/2 in 2+1 dimension.
Here we set the source operator in the origin as point source. 

\begin{figure}
\begin{center}
\includegraphics[width=90mm]{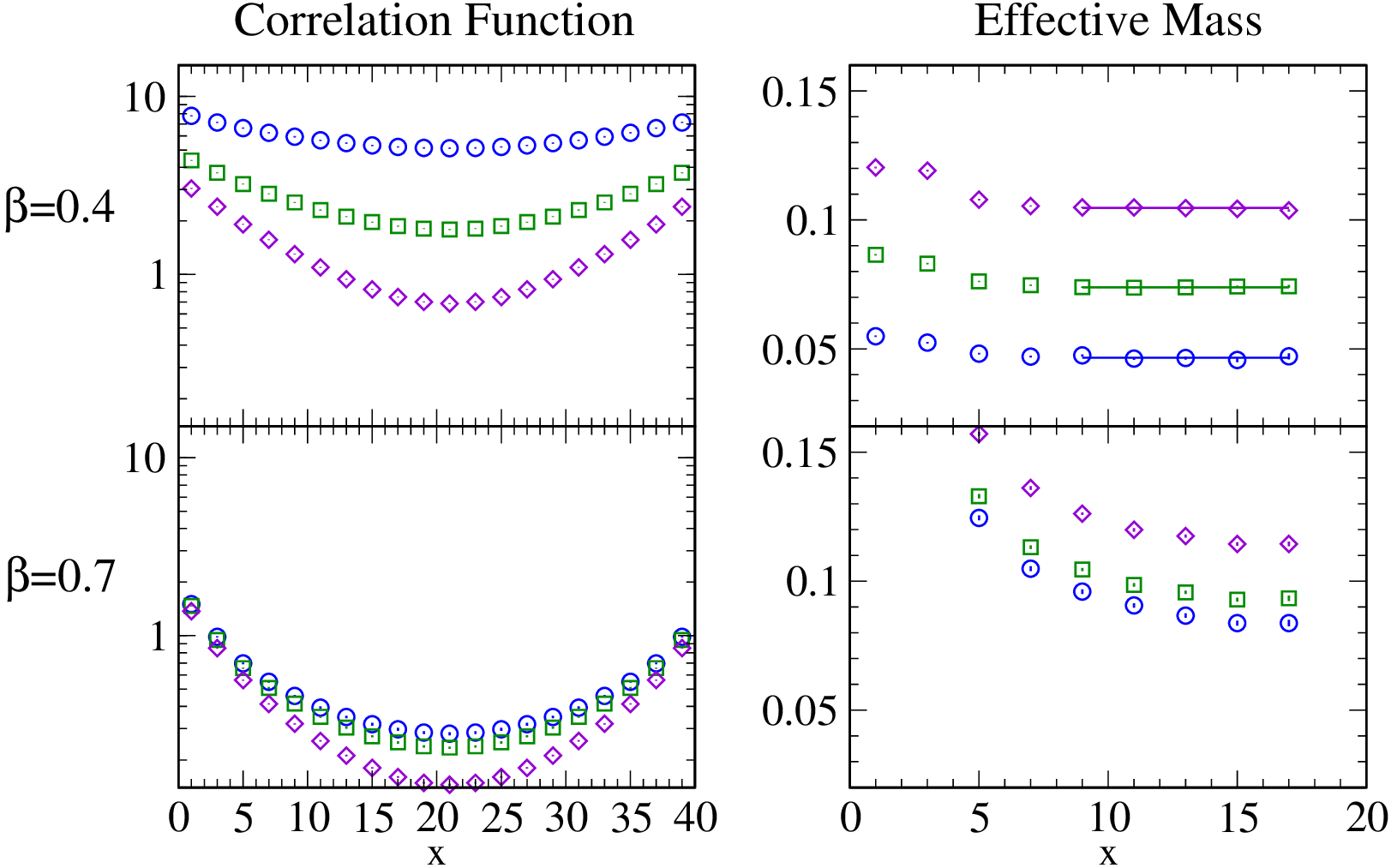}
\caption{$G_{\rm PS}(x)$ (left) and effective mass plot (right) at different 
$\beta$ in $M=0.001$(circle), $M=0.0025$(square) and $M=0.005$(diamond).
The straight lines are fitted result.}
\label{fig:picorr}
\end{center}
\end{figure}

Figure \ref{fig:picorr} shows the clear shape of exponential function 
of $G_{\rm PS}$, and effective mass plot explicitly illustrates 
that in $\alpha > \alpha_c$ there is plateau in $x\ge 9$, 
however in $\alpha < \alpha_c$ plateau is not observed.
This result indicates that NG boson state following in Eq.(\ref{eq:G_PS})
appears as a consequence of spontaneous $\chi$SB in $\alpha > \alpha_c$. 

\begin{figure}
\begin{center}
\includegraphics[width=70mm]{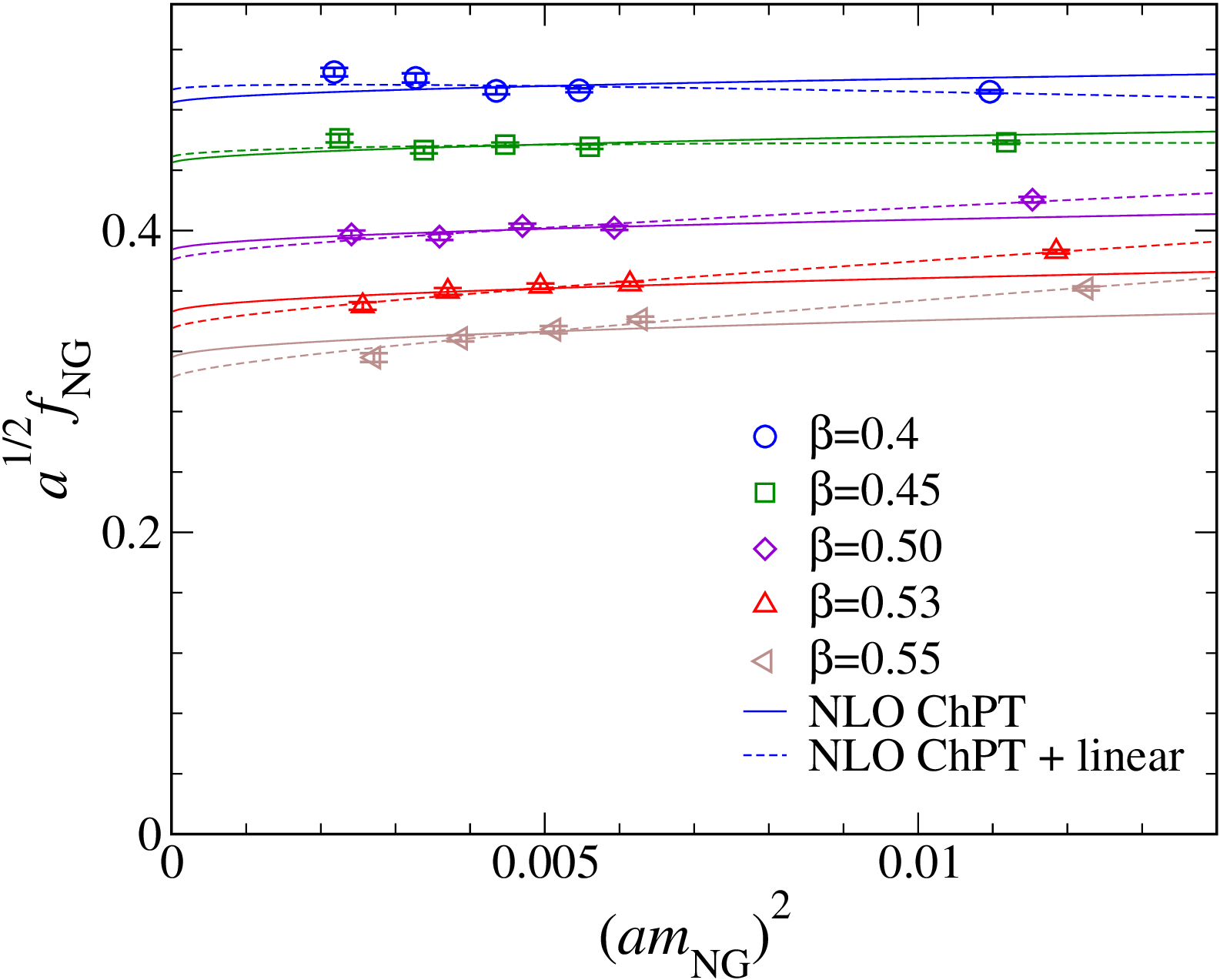}
\hspace{3mm}
\includegraphics[width=74mm]{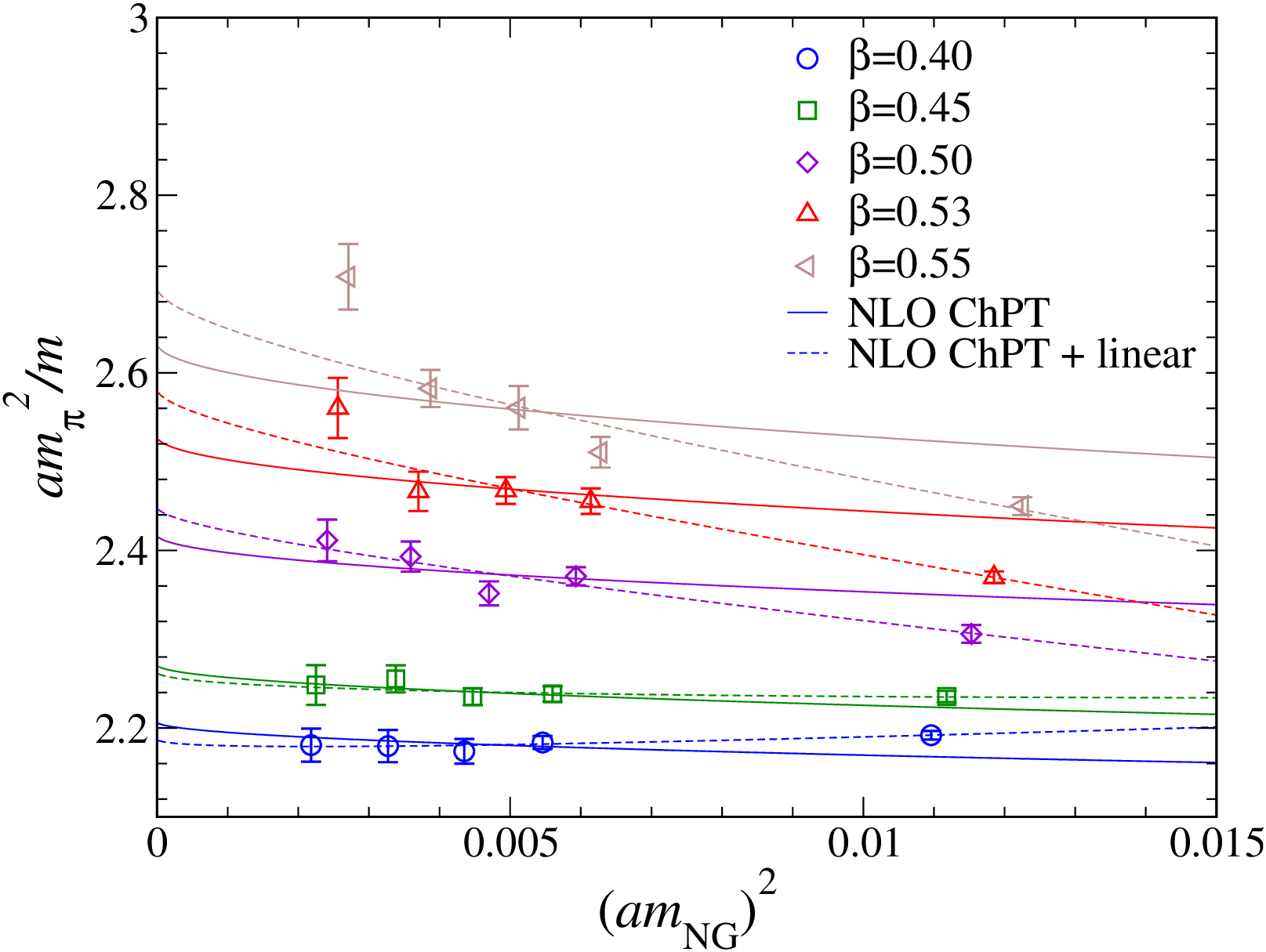}
\caption{$M$ dependence of $f_{\rm NG}$ (left) and $m_\pi^2/M$ (right). 
Solid line denote the LO ChPT in the range of $0.001\le m_{\rm NG}^2\le 0.0025$
and dashed one denotes the fitting function 
in LO ChPT plus linear term in the range of $0.001\le m_{\rm NG}^2\le 0.005$.}
\label{fig:fpi_mpi}
\end{center}
\end{figure}

Using the value obtained from fitting $G_{\rm PS}$ with function
in Eq.(\ref{eq:G_PS}), we show the lattice results of
$f_{\rm NG}$ and $m_{\rm NG}^2/M$ in each NG boson mass squared in 
Figure \ref{fig:fpi_mpi}.
To evaluate the chiral condensate from these results we can 
use the GMOR relation; $\Sigma = f_0 m_\pi^2/(2M)$ as in the 
case of QCD where $f_0$ is a value of $f_{\rm NG}$ in $M=0$.
According to the loop calculation in chiral perturbation theory (ChPT)
in 3-dimension, we set the fitting function
\begin{equation}
f_{\rm NG} = f_0(1+m_{\rm NG}/(4\pi f_0^2)) + cm_{\rm NG}^2,\quad
m_\pi^2/M = 2B(1-m_{\rm NG}/(4\pi f_0^2)) + dm_{\rm NG}^2,
\label{eq:fpi_mpi}
\end{equation}
where $f_0$, $B$, $c$ and $d$ are fitting variables.
We introduce $c$ and $d$ as correction terms to higher order 
effect than LO ChPT.
We attempt to compare the extrapolated result with 
different fitting range of $m_\pi^2$, and 
including the data at $M=0.005$ to extrapolate $M=0$ limit
we use a function in Eq.(\ref{eq:fpi_mpi}) besides
using data up to $M=0.0025$ we exclude $c$ and $d$ terms. 
In Figure \ref{fig:fpi_mpi} we see that in $\alpha > \alpha_c$ 
($\beta=0.4,\,0.45,\,0.5$) the chi-square fitting with both NLO ChPT 
and NLO CHPT + linear term works better than in $\alpha < \alpha_c$.
This is also consistent with picture that spontaneous $\chi$SB 
above $\alpha_c$ occurs in accompany with NG boson. 
In table \ref{tab:fpi_mpi} we show preliminary results of 
comparison of $\Sigma$ with three different ways in above $\alpha_c$. 
These are consistent results within 1--2$\sigma$ under a few \% accuracy. 

\begin{table}
\begin{center}
\caption{Fitting results of chiral condensate obtained by spectral 
function via Banks-Casher relation and ChPT with NLO and NLO+linear. 
We also describe the value of $\chi^2$/dof. 
The error is only statistical one.}
\label{tab:fpi_mpi}
\begin{tabular}{cccccc}
\hline\hline
$\beta$ & $\Sigma$ (spectral) & $\Sigma$ (NLO ChPT) & $\chi^2$/dof 
        & $\Sigma$ (NLO ChPT+linear) & $\chi^2$/dof \\
\hline
0.4  & 0.2433(48) & 0.2587(8) & 12.2(4.1) & 0.2657(15) & 6.1(2.9) \\
0.45 & 0.2140(45) & 0.2243(7) & 4.1(2.3) & 0.2275(13) & 2.7(1.9) \\
0.5  & 0.1694(51) & 0.1807(6) & 1.5(1.4) & 0.1766(13) & 2.9(2.0) \\
\hline\hline
\end{tabular}
\end{center}
\end{table}

\section{Summary and discussion}
We perform the {\it ab initio} calculation of QED model with 2+1 dimensional
fermion brane (bQED$_3$) in Monte-Carlo method. 
Due to preserving gauge symmetry
we take into account not only coupling constant but also
fermi-velocity as theoretical parameters. 
In this proceedings we numerically show the strong evidence of $\chi$SB 
and critical coupling constant from three kinds of way; 
chiral condensate, spectral density and NG boson spectrum. 
Although we fix the ``bare'' velocity parameter in 0.1 which 
is relatively larger than naive estimate ($v\sim O(10^{-3})$), 
we have the consistent result with expected in $\chi$SB phenomena.
This is feasible study to search the applicability to study of 
phase structure of Graphene.

The calculations were performed by using the RIKEN Integrated Cluster of Clusters 
(RICC) facility. 
This work is supported by the Grant-in-Aid of the Japanese Ministry of Education
(No. 20105002, 23105714(MEXT KAKENHI grant)).

\end{document}